\documentstyle[11pt,epsf]{article}
\def\beq{\begin{eqnarray}}
\def\eeq{\end{eqnarray}}
\def\bea{\begin{eqnarray*}}
\def\eea{\end{eqnarray*}}
\def\be{\begin{eqnarray}}
\def\ee{\end{eqnarray}}

\def\centeron#1#2{{\setbox0=\hbox{#1}\setbox1=\hbox{#2}\ifdim
\wd1>\wd0\kern.5\wd1\kern-.5\wd0\fi
\copy0\kern-.5\wd0\kern-.5\wd1\copy1\ifdim\wd0>\wd1
\kern.5\wd0\kern-.5\wd1\fi}}
\def\ltap{\;\centeron{\raise.35ex\hbox{$<$}}{\lower.65ex\hbox{$\sim$}}\;}
\def\gtap{\;\centeron{\raise.35ex\hbox{$>$}}{\lower.65ex\hbox{$\sim$}}\;}

\def\singleandthirdspaced{\baselineskip=\normalbaselineskip\multiply
    \baselineskip by 130\divide\baselineskip by 100}
\def\singlespaced{\baselineskip=\normalbaselineskip}

\newcommand{\newc}{\newcommand}
\newc{\qbar}{{\overline q}}
\newc{\Kahler}{K\"ahler }
\newc{\deltaGS}{\delta_{\rm GS}}
\begin{document}
\begin{titlepage}
\begin{flushright}
{\large hep-th/0303076 \\ SCIPP-2003/09 \\
MIFP-03-04\\ PUPT-2075\\
}
\end{flushright}

\vskip 1.2cm

\begin{center}

{\LARGE\bf Closed String Tachyons and Their Implications
for Non-Supersymmetric Strings}

\vskip 1.4cm

{\large  Michael Dine and Elie Gorbatov}
\\
\vskip 0.4cm
{\it Santa Cruz Institute for Particle Physics,
     Santa Cruz CA 95064  } \\
\vskip 0.4cm
{\large  Igor R. Klebanov\footnote{On leave from
Princeton University, Princeton, NJ 08544}
}
\\
\vskip 0.4cm
{\it School of Natural Science, Institute for Advanced Study,
Princeton, NJ 08540}\\
\bigskip
{\it and}\\
\bigskip
{\it George P. \& Cynthia W. Mitchell Institute for Fundamental
Physics}\\
{\it Texas A\& M University, College Station, TX 77843}\\

\vskip 0.4cm
{\large  Michael Krasnitz}\\
\vskip 0.4cm
{\it Joseph Henry Laboratories, Princeton University, Princeton, NJ 08544}\\

\vskip 1.5cm

\begin{abstract}
Closed string tachyons have long been somewhat mysterious.
We note that there is often a regime in the classical moduli space
in which one
can systematically compute the effective action for such
fields.  In this regime, the tachyon is light, and cannot be
integrated out.  Instead, one must consider the combined dynamics
of gravitons, moduli, tachyons and other light fields.  We compute
the action and find that the quartic term for the
tachyon is positive in the field definition where the
tachyon has no derivative coupling to the radion. We study
the evolution of isotropic, homogeneous configurations and find that
typically the system is driven to regions where the calculation is no longer
under control.

\end{abstract}

\end{center}

\vskip 1.0 cm

\end{titlepage}
\setcounter{footnote}{0} \setcounter{page}{2}
\setcounter{section}{0} \setcounter{subsection}{0}
\setcounter{subsubsection}{0}

\singleandthirdspaced

\section{Introduction}

Most known non-supersymmetric string theories contain tachyons.
For years there seemed to be no reliable technique in
string theory to compute the
potential for tachyonic fields (nor was it
clear that such a potential is meaningful), and so
it was hard to tell whether they are similar to familiar sorts
of field theory instabilities, or represent some new phenomena.
If the latter, one could imagine that they might even signal some
difficulty or inconsistency in these vacua,
which would be of great interest.

Luckily, for tachyons in open string theory, the questions above
have been largely resolved recently. In particular,
the decay of a D-brane anti D-brane pair in the superstring
\cite{Sen}, which contains open string tachyons when the
brane antibrane separation is smaller than a critical value \cite{BS},
has been
interpreted as tachyon condensation. A similar interpretation
applies to the decay of
an unstable D-brane of the bosonic string \cite{Senb,SZ}.
The off-shell open string tachyon potential can be calculated
using the techniques of open string field theory
\cite{KS}, and
it contains a local minimum corresponding to the state where
D-branes have decayed \cite{SZ}.

The fate of tachyons of closed string theory remains
largely mysterious, however. In particular, it is not clear
whether there exists a minimum of the tachyon potential
similar to the one found in the open string theory.
The difficulty in computing the tachyon potential is well known.
The potential is a zero-momentum quantity, while the tachyon
vertex operator is only defined on shell, far from zero momentum.
Closed string field theory is not known well-enough to carry
out the necessary off-shell calculations.

There is, however, one set of circumstances in which on can
compute the tachyon potential on shell.  Many non-supersymmetric theories
have, in their classical moduli spaces, critical points where
tachyons appear/disappear.  The simplest such case to study is
that due to Rohm \cite{rohm}; more extensive examples have been
enumerated in  \cite{ginspargvafa}.  In Rohm's model, Type II string
theory is compactified on a circle (torus), with boundary conditions
such that bosons are even but fermions odd in the extra dimension
(Scherk-Schwarz
boundary conditions). At large radius,
there is no tachyon.  As the radius decreases, however, a tachyon
appears at a critical radius, $R_o$.
At the critical radius, the ``tachyon
to be" is a massless particle, and it is straightforward to
compute its effective action.

The analogous situation in the open superstring arises when
a D-brane and an anti D-brane are at a critical separation
such that an open string connecting them is massless.
The effective potential for such modes was calculated in
\cite{Pesando} with the conclusion that the quartic term
is positive, in agreement with the idea that these
modes start to condense when the D-brane separation is reduced below
the critical one.

In this paper we carry out an analogous calculation for
the Rohm compactification of the closed string at the critical
radius, and find results similar to those found for the
open string.
At leading order in $\alpha^\prime$,
there is a quartic coupling of the tachyon, as well as a higher
derivative coupling to the radion.  With suitable field redefinitions,
the quartic term is positive. If $\Delta$ measures the distance to the
critical point (in the tachyonic direction), for small $\Delta$
these results remains correct to order $\Delta$.

Since, for small $\Delta$, the tachyon potential has a local
minimum,
it is tempting to then integrate out the tachyon, and
obtain a potential, already at the classical
level, for the moduli.  However, as we will see, this
is not consistent.  Near the critical
point, the moduli and the tachyon must be retained as
light fields.\footnote{A similar conclusion
was reached in studies of effective actions for Melvin backgrounds,
which are generalizations of the Rohm compactifications
\cite{Russ}. It was found, after a T-duality transformation,
the tachyon fits with other fields into a low-energy
effective action.}
Indeed, this is why the value of the
quartic coupling is
convention-dependent.  Field redefinitions can
render the coupling negative, while introducing various derivative
couplings to moduli.

In light of these comments, we focus on the evolution of homogeneous field
configurations of tachyons, moduli and gravitons
near the critical point.  This evolution is
independent of possible field redefinitions. Not surprisingly, we find
that the system is quickly driven out of the regime where the
effective action is calculable.  One is driven to values
of the radii where the tachyon is no longer nearly marginal; in addition,
one is driven to a regime where the theory is strongly coupled and the metric is
becoming singular.

The classical moduli, beyond the critical point, are not moduli at
all.  As one moves out of the region of controlled approximation,
the curvature of their potential is of order string scale.  In
this interior region of the ``moduli space," in other words, not
only is the system strongly coupled, but it is not clear what
degrees of freedom should be used to describe it.

Still, recent developments in string duality suggest some
plausible speculations.  Strong-weak coupling duality of
supersymmetric string theories suggests that in the very strongly
coupled region, the system becomes weakly coupled, and the
potential again tends to zero (from below).  So there is likely
some sort of minimum.  The AdS/CFT correspondence suggests that
this minimum is described by an isolated conformal 
field theory.\footnote{For related ideas on the end-point of
closed string tachyon condensation, see \cite{Barb}.}
One has no particular reason to think that this system is
supersymmetric, but one is free to speculate on this possibility.

What this analysis suggests is that the theory has some number of
stable, possibly non-supersymmetric, AdS states in its interior.  Actually,
AdS may, in some sense, be a misnomer, since there is no parameter
which makes a semiclassical, general relativistic analysis valid.
Perhaps one should
refer to these systems simply as conformal field theories, in some
strongly coupled domain.

\section{Rohm's Compactification}

Rohm proposed a simple model of supersymmetry breaking in string
theory \cite{rohm}.
Take Type II theory compactified on a circle, and impose
antiperiodic (Scherk-Schwarz) boundary conditions on fermions.  This breaks
supersymmetry.  It leads, at one loop, to a potential for the
moduli.  But for our purposes, what is more interesting is that,
while for large radius there is no tachyon, a tachyon does appear
at a finite value of the radius.  In the RNS formulation, for
example, the GSO projection is different in sectors with even and
odd winding.  Alternatively, in the Green-Schwarz formulation, the
Green-Schwarz fermions are antiperiodic in the odd winding number
sectors.  In general, for bosons, the momenta are
($\alpha^\prime = {1 \over 2}$)
\beq
k_L = {m \over 2R}-nR\ ,\qquad  k_R = {m \over 2R} +nR.
\eeq
Thus, for example, in the sector with winding number
one, the mass formula is ($\alpha^\prime = {1 \over 2}$):
\beq
L_0=  k_L^2 -1 = n^2 R^2 -1
\eeq
where in the last expression we have taken $m=0$.
So at $R=1$, a tachyon appears.  At smaller radius, more and more tachyonic
modes appear. The small radius limit of the theory
is in fact $T$-dual to a ten-dimensional
theory with a tachyon known as the type 0 string theory
\cite{DH}.

There may exist another interpretation of our results where the compact
dimension is the Euclidean time. This interpretation
applies to the thermal string theory, and the critical
radius corresponds to the Hagedorn phase transition \cite{Kog}--\cite{AW}.
Formally, the calculations which follow can be applied to the
system near the Hagedorn transition.  The interpretation of these
results, however, raises a number of subtle issues, which we leave
for future work \cite{joep}.

\section{The Effective Action}

At the critical point, the
would-be tachyon is massless, so we can calculate its effective
action
using on-shell vertex operators.  In the RNS formulation, for
example, the tachyon vertex operator in the minus one ghost number picture
is particularly simple \cite{KT}:
\beq
V_{-1} = e^{2i k_L X_L+2i k_R X_R}e^{i p \cdot x}e^{-\gamma}
\eeq
whereas in the ghost number zero picture it is
\beq
V_0 = (p \cdot \psi -2 k_L\psi_5)(p \cdot \tilde \psi - 2 k_R\tilde \psi_5)
 e^{2i k_L X_L+2i k_R X_R}e^{i p \cdot x}
 \eeq

Take the initial momenta, $(k_L,k_R,p)$ to be
$(1,-1,p_1),(-1,1,p_2)$ and the final momenta to be
$(-1,1,p_3),(1,-1,p_4)$ (all momenta are defined to flow into the
diagram).  Then the scattering amplitude is obtained from
\beq
<V_{-1}(k_1,p_1,z_1)V_{-1}(k_2,p_2,z_2)V_{0}(k_3,p_3,z_3)V_{0}(k_4,p_4,z_4)>
\eeq
$$
~~~~~={\vert z_1-z_4 \vert^2 \over \vert z_1-z_2 \vert^4}
{\vert z_2 -z_3 \vert^2 \over \vert z_2 -z_4 \vert^2 \vert z_3 -z_4 \vert^4
\vert z_1 -z_3 \vert^2} \Pi \vert z_i-z_j\vert^{-{p_i \cdot p_j \over 2}}
(-4+s/2)^2$$
where the last factor comes from the contraction of the
fermions, and $s, t, u$ are the Mandelstam variables:
$$ s =-(-p_1+p_2)^2, \; \; t = -(p_1+p_3)^2, \; \; u = -(p_1+p_4)^2.$$

Taking $z_1=\infty,z_2=0,z_3=1,z_4=z$, the amplitude is
\beq
{\cal A}= c\int d^2 z{(1+s/8)^2 \over \vert
1-z\vert^{4+s/4}\vert z \vert^{2+t/4}}
\label{lowenergy}
\eeq
$$ ~~~~=c^\prime {\Gamma(-t/8)\Gamma(-s/8) (1-u/8)\Gamma(1-u/8)
\over \Gamma(-1+u/8)\Gamma(1+t/8)\Gamma(1+s/8)}$$
$$~~~~~=c^{\prime}{u \over ts}(1-u/8)^2
 {\Gamma(1-t/8)\Gamma(1-s/8) \Gamma(1-u/8)
\over \Gamma(1+u/8)\Gamma(1+t/8)\Gamma(1+s/8)} $$
$$=c^{\prime}\left [{1 \over t}+{1 \over s} +({s \over 4t}+{t \over
4s})+1/2+ {s^2 \over 64 t} + {t^2 \over 64 s} + {3 s \over 64}
+ {3 t \over 64}+ \ldots \right ]\ .$$
In the last step, we have used energy-momentum conservation, $u=-t-s$;
the terms we
have dropped are suppressed by two powers of $s,t$.

In order to obtain the contact terms in the effective action, we need to subtract the
graviton, gauge boson and radion exchanges.
The graviton exchange (which we will use to normalize the string amplitude)
can be obtained from the effective action,
following closely the analysis of
\cite{KT}. Expanding the gravitational action,
we find
\be
\int d^9x{1\over 2}[h_{mn}(\Delta+...)h_{mn}+
{1\over 2}\phi \Delta \phi^*-{\cal T}_{mn} h_{mn}]
\ee
where
$$\Delta = -\partial^2, g_{mn}=\delta_{mn}+h_{mn},$$
and
\be
{\cal T}_{mn} = {1 \over 2}
[\partial_m \phi^*\partial_n \phi+ \partial_n \phi^*\partial_m \phi-
\delta_{mn}(\partial_k \phi^*) (\partial^k \phi)
]\ .
\ee
Integrating out the graviton to leading order,
we get the exchange amplitude which may be written as a
contribution to the S-matrix generating functional ${\cal S}(\phi)$
$${\cal S}(\phi )=\int d^9x[{1 \over 2}\phi \Delta \phi^*+W],$$
$$W = -{1 \over 2} {\cal T}_{mn}\Delta^{-1}_{mn,kl}{\cal T}_{kl},$$
where
$\Delta^{-1}_{mn,kl}=(\delta_{m(k}\delta_{l)n}-
{1 \over D-2}\delta_{mn}\delta_{kl})\Delta^{-1}$
is the graviton propagator in the harmonic gauge.
Evaluating the contractions, we find
\be
W = \int \Pi {d^9k_i \over (2\pi)^9}e^{ik_i \cdot x}
{\cal W}(k_1,k_2,k_3,k_4)
\phi (k_1)\phi (k_2)\phi^*(k_3)\phi^*(k_4),
\ee
$${\cal W} = {1 \over 32}[{s^2+u^2 \over t}+{u^2+t^2 \over s}-(t+s)]
\ .$$

The
non-derivative radion coupling to the tachyon can be determined
from the mass formula for the tachyon:  $m^2 = 4R^2-1$.  There also
may be derivative
couplings which are not visible in the on-shell 3-point function
for $\phi$, $\phi^*$ and the radion.
We can obtain both couplings by
examining the OPE of two tachyons directly.  This includes:
\beq
V_0(k_L,k_R,p) V_0(-k_L,-k_R,p^\prime) \sim
{4 k_L k_R \over \vert z-z^\prime \vert^2}
[k_L^2 + {p \cdot p^\prime \over
4}][k_R^2 + {p \cdot p^\prime \over
4}]\partial X_L \bar \partial X_R e^{i(p+p^\prime)\cdot x}+ \dots.
\label{ope}
\eeq
To get the gauge boson couplings, replace $2k_L \partial x_L$ by
$p_{\mu} \partial x^\mu_L$, and similarly for the right.

So radion exchange gives a contribution,
\beq \label{radoff}
{1 \over 2s}(1+s/8)^4 + {1 \over 2t}(1+t/8)^4
\ .
\eeq
Gauge boson exchanges give a contribution
$${t+s/2 \over s} + {s+t/2 \over t}.$$

We
can make these contributions manifest by rewriting the scattering amplitude
as
\beq
c^\prime[(1/t+1/2)+(1/s+1/2)
+({s \over 4t}+1/8)
+({t \over 4s}+1/8) - 1/4 \eeq
$$~~~~~~+ ({(s+u)^2 \over 32 t} + {(t+u)^2 \over 32 s}-{(t+s) \over 32}+
{3s + 3t \over 128} )]
=c'[{\rm radion}+{\rm gauge~boson} - 3/4 +{3s + 3t \over 128}].
$$
The $-3/4$ represents a non-derivative quartic interaction;
the last term is a quartic interaction with
two derivatives.
To determine the sign of the quartic term, one just has to compare with the
expected sign from, e.g. the radion exchange.  A positive quartic coupling
should give a contribution of the same sign as the pole term;
so we have found a negative quartic coupling.

\section{Field Redefinitions}

The effective action is only defined up to field redefinitions.
The operator product
expansion  of eqn. (\ref{ope}) suggests that one take the coupling
\beq
\partial^2 R \vert \phi \vert^2 + {3 \over 8}
\vert \phi \vert^4 + {\rm four~derivative~terms}.
\eeq
In this case, the quartic coupling is negative.  Alternatively, one can
make a field redefinition,
\beq
R \rightarrow R - {1\over 2} \vert \phi \vert^2
\eeq
This yields
\beq
{\cal L}_{eff}= {1 \over 16}(\phi^* \phi)^2
+ {1\over 64}\phi^* \phi  {\partial_{\mu} \phi^*  \partial^\mu \phi}.
\eeq
This effective action, along with radion, gauge boson, and graviton exchanges
reproduces the amplitude of eqn. (\ref{lowenergy}).
In this form, the quartic
coupling is positive.

This form of the action could be derived directly had we taken the
radion exchange contribution of the form
\beq
{1\over 2s} + {1\over 2t}
\eeq
rather than (\ref{radoff}). Then the amplitude with all massless exchanges
subtracted is
\beq
{1\over 4} + {s+t\over 32}
\eeq
corresponding to ${\cal L}_{eff}$.

It is clear from this discussion that the sign of
the quartic coupling is ambiguous.
However, the low energy dynamics is not.
We will discuss these issues in section 6.

\section{Another Model}

There are other models to which one can apply this sort of analysis.
One example is provided by
the Rohm compactification, not for the Type II
theory, but for the heterotic theory (for definiteness, we can
discuss the O(32) theory; it is convenient to use the fermionic
formulation for the left movers).

The analysis is quite similar to the Type II case.  There is again
a tachyon at the (same) critical radius, in the vector
representation of $SO(32)$.  In the
ghost number $-1$ picture, we can take the vertex operator to be
\beq
V_{-1} = \lambda_a e^{i k_L X_L +i k_R X_R + i p \cdot x}
e^{-\gamma}.
\eeq
We can now make the computation of the four point function very
similar to that of the Type II case, if we judiciously choose the
quantum numbers of the tachyons.
One can take, say, the two operators in the $-1$ picture to
carry gauge index $1$, and the two in the zero picture to
carry gauge index $2$.  Then the calculation of the amplitude is
identical to that in the Type II case, except that the factor
$(1+s/8)^2$ is now $(1+s/8)$.  In extracting the quartic coupling,
however, we need to note that the massless exchanges are
different.  First, the couplings of the radial mode are now
proportional to $(1+s/8)$, etc.  The quantum numbers of the
exchanged particles are also different.  In the $t$-channel, for
example, the exchanged particle carries $O(32)$ quantum numbers;
it is a dimensionally reduced gauge boson.  Similarly, the
quantum numbers of the exchanged gauge bosons in the two channels
are different.
One again obtains derivative couplings of the radion to the tachyon, and a negative
quartic coupling.  A suitable field redefinition eliminates the
derivative couplings, and yields a positive quartic coupling.

\section{Tachyon and Moduli Dynamics; Speculations on the ``Moduli
Potential"}

We have seen in two examples that the question
of whether the tachyon potential has a calculable minimum
is not unambiguous.
Field redefinitions allow us to absorb part of the coefficient of the quartic
coupling into operators such as $\partial^2 R |\phi|^2$.
The low energy physics
should be invariant under these redefinitions.
If the quartic potential were positive, one might have hoped to
find the minimum of the tachyon potential and integrate it out.
This would leave the puzzle of field redefinitions.  In fact,
integrating out the tachyon in this way, at least in the regime
where we know how to calculate the tachyon effective action, is
not consistent.

To see this, suppose that the tachyon potential has a minimum.   If for the
moment we call $R^2 = 1-\sigma$, then minimizing with respect
to the tachyon, the potential
behaves as
\beq
V= -{\sigma^2 \over g^2}.\eeq
This is a potential for the radion and the dilaton.
But the effective mass for the $\sigma$ field is of order one.
Before proceeding further, one observation about this potential is
in order.  It might seem that this potential becomes arbitrarily
negative for weak coupling.  But there is no simple meaning to
energy in general relativity, and to interpret this result it is
helpful to go to the Einstein frame.  There, in any dimension, the
potential depends on $g$ with a positive power.  So if there is a
minimum, or even a regime where the potential is negative, the
system is driven to strong coupling, where any analysis is likely
to break down.

In any case, given our result that the tachyon
cannot consistently be integrated out
near the critical point, it is necessary to keep both the fields
$\sigma$ and $\phi$ (as
well as the dilaton) in the low energy effective action in this
region of field space.
We have solved for the evolution of
homogeneous and isotropic field configurations with the equations
following from this action.  These equations are manifestly
covariant under field
redefinitions.  We have studied a variety of possible initial
conditions.  For example, we have taken the initial radius
such that the tachyon has a small negative mass-squared,
and started the tachyon near the minimum of the effective
potential with the positive quartic coupling.  Not surprisingly, the system is driven, with
typical initial conditions, to a regime in which the tachyon is no
longer approximately marginal (and in which there are, in fact,
more tachyons) and in which the coupling is becoming strong.

To summarize:  at this stage,
one has generated, {\it
classically}, a potential on the moduli space.  In other words,
the moduli are not moduli past the critical point.  It is
necessary, near the critical point, to study the dynamics of the
original moduli and tachyon together.  One is quickly driven,
however, to a regime where one does not have control of the
calculation.

We can speculate on the form of the moduli potential in the
strongly coupled region.  If we suppose that there is an
S-duality, then we might expect that there is a minimum in this
region.  It might be described by some suitable conformal field
theory.  There is no obvious small parameter, so this minimum will
occur when all scales are comparable and all couplings of order
one.

The alternative is
that the potential is unbounded
below.  This would be an exciting possibility, and would establish
that these vacua are not sensible, but it seems unlikely.  It is true that
at the minimum, if it exists, the appropriate degrees of freedom
to describe the system are not the radion and dilaton, but
others.  An isolated CFT is probably a good model for this.

\noindent
{\bf Acknowledgements:}

\noindent
We are grateful to T. Banks, A. Tseytlin and E. Witten for useful discussions.
The work of M.D. and E.G.
was supported in part by the U.S.
Department of Energy.  That of I.K. and M.K.
was supported in part by the National Science
Foundation under Grant No. PHY-9802484.
Any opinions, findings, and conclusions or recommendations expressed in
this material are those of the authors and do not necessarily reflect
the views of the National Science Foundation.


\end{document}